\documentclass[twocolumn,english,prb,showpacs,superscriptaddress,floatfix]{revtex4-1}
\usepackage[T1]{fontenc}
\usepackage[latin9]{inputenc}
\usepackage{babel}
\usepackage{amsmath}
\usepackage{amssymb}
\usepackage{graphicx}
\usepackage[colorlinks=true,citecolor=blue,linkcolor=blue,urlcolor=blue,breaklinks=true]{hyperref}

\makeatletter
\@ifundefined{textcolor}{}
{%
 \definecolor{BLACK}{gray}{0}
 \definecolor{WHITE}{gray}{1}
 \definecolor{RED}{rgb}{1,0,0}
 \definecolor{GREEN}{rgb}{0,1,0}
 \definecolor{BLUE}{rgb}{0,0,1}
 \definecolor{CYAN}{cmyk}{1,0,0,0}
 \definecolor{MAGENTA}{cmyk}{0,1,0,0}
 \definecolor{YELLOW}{cmyk}{0,0,1,0}
}

\usepackage{braket}

\makeatother

\begin{document}

\title{Fate of the cluster state on the square lattice in a magnetic field}

\author{Henning Kalis}

\email{henning.kalis@tu-dortmund.de}

\affiliation{Lehrstuhl für Theoretische Physik I, Otto-Hahn-Straße 4, TU Dortmund,
D-44221 Dortmund, Germany}

\author{Daniel Klagges}

\email{daniel.klagges@tu-dortmund.de}

\affiliation{Lehrstuhl für Theoretische Physik I, Otto-Hahn-Straße 4, TU Dortmund,
D-44221 Dortmund, Germany}

\author{Rom\'an Or\'us}

\email{roman.orus@mpq.mpg.de}

\affiliation{Max Planck Institute of Quantum Optics, Hans-Kopfermann-Strasse 1, 85748 Garching, Germany}

\author{Kai Phillip Schmidt}

\email{kai.schmidt@tu-dortmund.de}

\affiliation{Lehrstuhl für Theoretische Physik I, Otto-Hahn-Straße 4, TU Dortmund,
D-44221 Dortmund, Germany}
\begin{abstract}
The cluster state represents a highly entangled state which is one central
object for measurement-based quantum computing. Here we study
the robustness of the cluster state on the two-dimensional square lattice
at zero temperature in the presence of external magnetic fields
by means of different types of high-order series expansions and variational techniques using infinite Projected Entangled Pair States (iPEPS). The phase diagram displays a first-order phase transition line ending in two critical end points. Furthermore, it contains a characteristic self-dual line in parameter space allowing many precise statements. The self-duality is shown to exist on any lattice topology. 
\end{abstract}

\pacs{03.67.-a,03.67.Lx,75.40.-s,}

\maketitle
\section{Motivation}
\label{Sect:Motivation}

The exploitation of quantum mechanics to build a quantum computer is a very active area in current research, because it is expected to be capable of solving classically hard problems in a polynomial amount of time \cite{Deutsch08071985} yielding a deeper understanding of the quantum world. To this end it has been shown that a universal quantum computer can be built by only a small set of elementary operations, namely arbitrary single-qubit rotations plus certain two-qubit gates like CZ or cNOT\cite{PhysRevA.52.3457,doi:10.1139/P08-112}. Especially the two-qubit operations turn out to be complicated to implement in experiment. 

Measurement-based quantum computing is a fascinating alternative approach for a quantum computer\cite{PhysRevLett.86.5188}. The essential idea is to prepare a highly-entangled initial quantum state on which only single-qubit measurement are sufficient to run a quantum algorithm. Meaurements with respect to an arbitrary basis are easy to perform in experiment. This feature comes with the price, that the initial state is hard to prepare in nature. One class of such highly-entangled states useful for measurement-based quantum computation are cluster states. 

One natural way of realizing a cluster state would be to cool down appropriate Hamiltonians having the cluster state as a ground state. Indeed, so-called cluster Hamiltonians exist but contain typically multi-site interactions which are very rare in nature. As a consequence, simpler models containing solely two-spin interactions have been proposed in the literature having the cluster Hamiltonian as an effective low-energy model. But it has been shown recently that it is very challenging to protect approximative cluster states against additional perturbations\cite{arXiv:1111.5945v1}. Another approach to study such systems efficiently, could be to prepare the cluster Hamiltonian with a quantum simulator\cite{0034-4885-75-2-024401,Buluta02102009,citeulike:6932721}. However simulating multi-spin interactions with respect to the desired topology will probably be a challenge. 

In any case it is important to check whether the cluster state is stable and protected against additional perturbations. This has been the subject of several works in recent years which mostly concentrate on additional magnetic fields as a perturbation \cite{PhysRevA.80.022316,PhysRevLett.103.020506}. The latter studies either investigated the change of entanglement of the perturbed cluster state or explored the complete breakdown of the cluster state due to a phase transition which serves as an upper bound for measuremen-based quantum computing. 

Most is known for the perturbed one-dimensional cluster-state Hamiltonian, e.g. the case of a single magnetic field in $x$-direction can be solved exactly by fermionization giving a second-order phase transition \cite{PhysRevLett.103.020506}. Recently also a transverse Ising perturbation has been investigated \cite{0295-5075-95-5-50001}. But of special interest are two-dimensional lattice topologies for which a universal measurement-based quantum computer can be formulated \cite{2009arXiv0910.1116B}. 

In this work we concentrate on the perturbed cluster-state Hamiltonian on the square lattice. The case of an additional magnetic field in $z$- or $x$-direction has been already studied \cite{PhysRevA.80.022316}. However, a combination of both fields has never been investigated. This more complicated problem is the major topic of this work. The central aim from a solid-state perspective is therefore to obtain the phase diagram. This is achieved by combining analytical and numerical means. To be concrete, we will show analytically that the phase diagram contains a self-dual line in parameter space. A combination of high-order series expansion and variational techniques are then used to determine numerically the full phase diagram. Our main finding is the existence of a first-order phase transition line ending in two critical end points. Furthermore, we introduce a quasi-particle picture for the elementary excitations within the cluster-phase. Finally we investigate the fidelity of the cluster-state with the perturbed ground states depending on the strengh of the perturbation, further confirming the obtained phase diagram. Rigorously we then determine the threshold for the usablity for measurement based quantum computing with these states. 

The paper is organized as follows. In Sect.~\ref{Sect:Model} we introduce the cluster-state Hamiltonian in a magnetic field and we discuss certain limiting cases. Afterwards, we proof the existence of a self-dual line in parameter space. The numerical methods are introduced in Sect.~\ref{Sect:Methods} and the resulting phase diagram is presented in Sect.~\ref{Sect:PD}. The consequences for the usability in measurement-based quantum computing are discussed in Sect.~\ref{Sect:F}. Finally, in Sect.~\ref{Sect:Conclusion} the major findings of this work are discussed and embedded in possible future lines of research.  

\section{Model}
\label{Sect:Model}
\begin{figure}
\begin{centering}
\includegraphics[width=0.8\columnwidth]{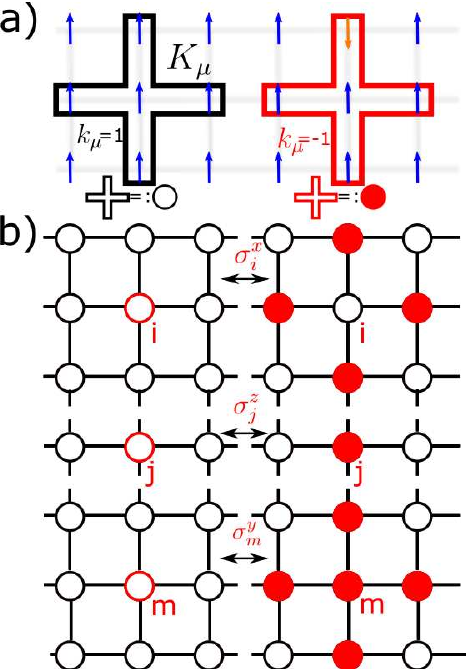} 
\par\end{centering}
\caption{(color online) (a) Eigenvalues $k_\mu$ of stabilizer operator $K_\mu$ on the square lattice with spin $1/2$ degrees of freedom on each lattice site. Since $K_\mu$ commutes with each of the star-like five spin interaction forms a superlattice, that again is defined on a square lattice. (b) Effect of $\sigma^\alpha$ operators on the effective superlattice. In conclusion the mainly investigated perturbation with a combined $h_x$,$h_z$-field can create or destroy (depending on the intial state) either $0,1,2$ or $4$ clusterons.}
\label{fig:overview}
\end{figure}

The cluster Hamiltonian introduced by Raussendorf and Briegel\cite{PhysRevLett.86.5188} has the cluster state as its unique ground state. The model we investigate is defined on the two-dimensional square lattice at zero temperature where there is a spin $1/2$ degree of freedom at each site of the lattice. The Hamiltonian reads
\begin{eqnarray}
\mathcal{H}_{\text{CL}} = -J\sum_{\mu}\sigma^x_{\mu}\displaystyle \bigotimes_{j \in \Gamma(\mu)} \sigma^z_j = -J\sum_{\mu}K_\mu \label{Mod:hcl}\text{,}
\end{eqnarray}
where $\Gamma(\mu)$ denotes the four nearest-neighbour spins of lattice site $\mu$ and the $\sigma^\alpha$ are the usual Pauli matrices with $\alpha\in\{x,y,z\}$. The cluster Hamiltonian $\mathcal{H}_{\text{CL}}$ is exactly solvable. This is a consequence of the large number of conserved quantities originating from the fact that all operators $K_\mu$ commute with each other and therefore with the full cluster Hamiltonian $\mathcal{H}_{\text{CL}}$. The conserved eigenvalue $k_\mu$ of each operator $K_\mu$ takes values $\pm 1$ (illustrated in Fig.~\ref{fig:overview}(a)), what is a direct consequence of $K_\mu^2 = 1$.

Consequently, the cluster Hamiltonian $\mathcal{H}_{\text{CL}}$ has an equidistant spectrum and its ground state for positive $J$ corresponds to the unique state having $k_\mu=1$ for all $\mu$, i.e. the energy per site is $-J$. Elementary excitations of the system are called clusterons. A one-clusteron state is defined by the flip of one eigenvalue $k_\mu$ to $-1$, i.e. the excitation gap for the creation of a single clusteron is $2J$. Clusterons of the Hamiltonian $\mathcal{H}_{\rm CL}$ are hardcore bosons being static and non-interacting. Let us remark that the cluster state is not topologically ordered in contrast to the so-called toric code \cite{Kitaev2003} which is also a stabilizer code but displays topological order. As a consequence, elementary excitations of the cluster Hamiltonian are not fractionalized and a single clusteron is a well defined excitation.\\
Here we are interested in the fate of the cluster state under an external uniform magnetic field. The full Hamiltonian of interest reads
\begin{eqnarray}
\mathcal{H} = \mathcal{H}_{\text{CL}} - \sum_{i,¸\alpha}\,h_\alpha^{\phantom{\alpha}}\, \sigma^\alpha_i \quad ,
\label{hamcl}
\end{eqnarray}
where the sum runs over all lattice sites $i$ and over all field directions $\alpha\in\{x,y,z\}$. A finite external magnetic field typically destroys the exact solvability of the model and one is confronted with a complicated two-dimensional many-body problem. Clearly, at very large external fields the ground state of the system corresponds to a polarized phase where all spins point in field direction. The highly-entangled cluster state will therefore be destabilized under an external field either by reducing the entanglement adiabatically or more drastically by a quantum phase transition, i.e. a macroscopic rearrangement of the ground state. 

We start to analyse $\mathcal{H}$ by discussing certain limiting cases where rather strong statements can be done: 

{\emph{--single $h_z$-field--}} The simplest case is the one where only the $h_z$-field is finite. In this limit the system remains exactly solvable. The latter is a consequence of the fact that the action of $\sigma^z_i$ on any eigenstate of $\mathcal{H}_{\text{CL}}$ characterized by the eigenvalues $k_\mu$ only flips the value of $k_i$, i.e. the operator $\sigma^z_i$ creates or destroys a single clusteron on site $i$. This process (and the effect of all Pauli matrices $\sigma^\alpha_i$) is illustrated in Fig.~\ref{fig:overview}(b). Effectively, one has a collection of $N$ independent two-level systems which can be solved analytically. The clusteron remains static and non-interacting. Its excitation energy is given by
\begin{eqnarray}
 \Delta = 2\sqrt{ J^2 + h_z^2 } \quad \text{.}
\end{eqnarray}
The one-clusteron gap $\Delta$ increases with increasing field. The cluster phase becomes more stable and it is adiabatically connected to the polarized high-field phase. In contrast, the entanglement of the ground state is strongly reduced and the usability for measurement-based quantum computing is lost for a finite value of $h_z$ \cite{PhysRevA.80.022316} (see also Sect.~\ref{Sect:F}).

{\emph{--single $h_x$-field--}} Next we focus on the case where only the $h_x$-field is finite. The cluster Hamiltonian in the presence of an $h_x$-field is not exactly solvable anymore. The action of $\sigma^x_i$ on a zero-field eigenstate is to flip all the four eigenvalues $k_j$ with site $j$ being a nearest neighbor of site $i$ (see Fig~\ref{fig:overview}). One is therefore left with a complicated many-body problem of interacting and mobile clusterons. Fortunately, the $h_x$-only case can be mapped to different models discussed in the literature in recent years. To be specific, the cluster Hamiltonian in the presence of an $h_x$-field is isospectral to the Xu-Moore model\cite{PhysRevLett.103.020506,PhysRevLett.93.047003} which is known to be isospectral to the quantum compass model\cite{0038-5670-25-4-R03} and to the toric code in a transverse field\cite{PhysRevB.80.081104}. Note that only the spectrum is the same for all models but not the degeneracies. As a consequence of the isospectrality of all four models, one can conclude that the cluster Hamiltonian in the presence of an $h_x$-field is self-dual and a first-order phase transition exists at $J=h_x$ separating the cluster phase and the polarized phase \cite{PhysRevB.78.064402,PhysRevB.80.081104}. To proof the self-duality we make use of the controlled-Z (CZ) transformation on every bond of the lattice. Reminding the matrix representation
\begin{equation}
[\text{CZ}] = \left(\begin{array}{cccc}
1 & 0 & 0 & 0       \\
0 & 1 & 0 & 0       \\
0 & 0 & 1 & 0       \\
0 & 0 & 0 & -1
\end{array}\right)
\end{equation}
one directly can confirm its unitarity. The application of the CZ will result in
\begin{eqnarray}
[\text{CZ}]\mathcal{H}(J,h_x)[\text{CZ}] = \mathcal{H}(h_x,J)\quad ,
\end{eqnarray}
what is effectively the exchange of $J$ and $h_x$.

{\emph{--single $h_y$-field--}} Interestingly $\mathcal{H}(J,h_y)$ also reveals to be self-dual. We will proof this via the application of the CZ and subsequent rotations in the Pauli basis that could be visualized on the Bloch sphere. Performing a CZ on every bond of the lattice will lead to the expression
\begin{equation}
[\text{CZ}]\mathcal{H}(J,h_y)[\text{CZ}] = -J\sum_i\sigma_i^x - h_y\sum_\mu\sigma^y_{\mu}\displaystyle \bigotimes_{j \in \Gamma(\mu)} \sigma^z_j\quad .
\end{equation}
Applying a $\pi$ rotation around the $y$-direction and a $\pi$/$2$ rotation around the $z$-axis one transforms the Hamiltonian in the original one with the desired exchange of the parameters $J$ and $h_y$.\\
Once more than one field $h_x$, $h_y$, or $h_z$ is finite, no rigorous results are known in the literature. This is the main motivation of this work. In the following we will concentrate on the case $h_y=0$. Interestingly, the model displays a self-dual line for this case. One can show again that performing a unitary transformation with the operator CZ on every bond of the lattice, will lead to the expression
\begin{eqnarray}
\text{[CZ]}\mathcal{H}(J,h_z,h_x)\text{[CZ]} = \mathcal{H}(h_x,h_z,J)\quad .
\end{eqnarray}
This is a direct consequence of $[CZ,\sigma^z\otimes 1] = 0$. So one finds one self-dual point at $J=h_x$ for each value of $h_z$ which reduces to the self-duality discussed above for the $h_x$-only case if $h_z=0$. This analytical property will strongly help us to numerically analyze Hamiltonian $\mathcal{H}$. Another remarkable aspect of the CZ transformation is its independence of the lattice topology. Therefore the self duality holds also on other lattices where one can define the cluster Hamiltonian on neighbouring bonds.

Our main goals are the following. First, we would like to deduce the full zero-temperature phase diagram. We are therefore interested in phase transitions from the highly-entangled cluster phase into polarized phases present at large fields. Such phase boundaries of the cluster phase certainly represent upper bounds for the usability of measurement-based quantum computing. Second, we will analyse the fidelity of the ground state in the presence of external fields with the exact cluster state at zero field. This allows more accurate conclusions about the robustness of the cluster state under the presence of external perturbations and its practical usefulness.

\section{Methods}
\label{Sect:Methods}

We study Hamiltonian $\mathcal{H}$ by combining high-order series expansions and variational calculations using infinite projected entangled pair-states (iPEPS) \cite{2004cond.mat..7066V}. In the following we will introduce the most important properties of both individual methods. Afterwards, we describe how to combine both techniques.

\subsection{Series expansions}

Our aim is to calculate a high-order series expansion of the ground-state energy per site, of the one-particle gap, and of the fidelity per site for different limits of the perturbed cluster Hamiltonian. We have used a partitioning technique provided by L\"owdin \cite{Loewdin:969,yao:278} to calculate the energetic properties of the system. The fidelity has been calculated \cite{arXiv:1111.5945v1} by a projector method introduced by Takahashi \cite{0022-3719-10-8-031}. A more detailed discussion of the fidelity and its calculation is presented in Sect.~\ref{Sect:F}. Here we concentrate on describing the most relevant properties of L\"owdin's approach giving us the essential quantities to determine the phase diagram.   

The L\"owdin approach is highly related to the Rayleigh-Schr\"odinger and Brillouin-Wigner perturbation theory but provides faster convergence in the case of degeneracies \cite{yao:278}. In essence, all approaches try to find an approximative solution of the eigenvalue problem
\begin{eqnarray}
\mathcal{H}|\psi_n\rangle = E_n|\psi_n\rangle \text{,}\label{hpsiepsi}
\end{eqnarray}
where usually the Hamiltonian $\mathcal{H} = \mathcal{H}_0 + \lambda V$ is split into an unperturbed part $\mathcal{H}_0$ and a perturbation $V$ which is adiabatically turned on with the real parameter $\lambda$. Consider the $n$-th eigenvalue of $\mathcal{H}_0$ to be $g$-fold degenerate.  Using the partitioning technique we define two projection operators $P$ and $Q$ as follows:
\begin{equation}
P = \sum_{j = 1}^{g} | \psi^{(0)}_{n,j}\rangle\langle \psi^{(0)}_{n,j}| \ \ \ \ \ \ Q = 1 - P  \text{,}
\end{equation}
where $|\psi^{(0)}_{n,j}\rangle$ denote the unperturbed eigenstates of $\mathcal{H}_0$. Therefore $P$ is a projection operator on the unperturbed eigenspace of $\mathcal{H}_0$ and $Q$ is the projection operator on the complementary (orthogonal) space. Furthermore, let us define $R := (E_n^{(0)} - \mathcal{H}_0)^{-1}$ to be the resolvent where $E_n^{(0)}$ corresponds to the unperturbed eigenenergy of the states $|\psi^{(0)}_{n}\rangle$. In the following we denote by $E_n^{(j)}$ with $j\in\{0,1,2,\ldots\}$ the energy correction in order $j$ perturbation theory of the eigenvalue problem Eq.~\eqref{hpsiepsi}. In fact, in the non-degenerate case ($g = 1$) the application of the characteristc L\"owdin operator sequence in order $j$
\begin{eqnarray}\label{Eq:Loewdin}
\mathcal{O}^{(j)} = PV\sum_{m=0}^{j-1}\left[\sum_{i=0}^{m-1}\left(-R\sum_{k=1}^{i} E_n^{(k)}\right)^iRQV\right]^{m}P
\end{eqnarray}
corresponds to the energy correction terms of the Rayleigh-Schr\"odinger perturbation theory. Calculating the expectation value for the ground-state energy in order $j$ then can be achieved by computing
\begin{eqnarray}
E_0^{(j)} = \langle\psi_0^{(0)}|\mathcal{O}^{(j)}|\psi_0^{(0)}\rangle\quad\text{.}
\end{eqnarray}
Therefore the L\"owdin method provides perturbative corrections of expectation values for an observable up to a desired order $j$. We have calculated the general operator sequence given in Eq.~\eqref{Eq:Loewdin} up to order 16. Additionally, we have reached order 20 for the simpler case where the first-order contribution given on the operator level by $PVP$ vanishes. Since we are interested in the phase diagram, which in general can contain first- and second-order phase transitions, it is necessary to determine the ground-state energy per site $e_0 \equiv E_0/N$ and the one-particle gap $\Delta$ as a high-order series expansion. 

Let us stress that $E^{(0)}_1$ is $N$-fold degenerate in all perturbative limits considered in this work. In the high-field limit the unperturbed first excited states $E^{(0)}_1$ correspond to spin flip excitations in the polarized phases and in the cluster phase $E^{(0)}_1$ refers to the bare one-clusteron energy. 

Unfortunately, it is not possible to calculate the gap momentum with L\"owdins method because one has no access to hopping elements. We therefore have also used the method of Takahashi \cite{0022-3719-10-8-031} and perturbative continuous unitary transformations \cite{ANDP:ANDP19945060203,springerlink:10.1007/s100510050026,0305-4470-36-29-302} allowing a high-order series expansion of the hopping elements. With these methods one usually calculates the one-particle dispersion $\omega(k_x,k_y)$. The one-particle gap is then identified as the global minimum obtained for a certain momentum $\mathbf{k}_{\rm{min}}$. Once the gap momentum is identified, one can again use L\"owdin's approach by constructing an eigenstate of the system having this specific momentum (see below).

In order to compute the dispersion $\omega(k_x,k_y)$ one has to find all possible hopping elements $a^{(j)}_{(l,m)}$ in a given order $j$. A hopping element from site $(x,y)$ to site $(x+l,y+m)$ on the square lattice is defined as
\begin{eqnarray}
a^{(j)}_{(l,m)} = \langle\psi_1^{(0)}|_{(x+l,y+m)}\mathcal{H}^{(j)}_{\text{eff}}|\psi_1^{(0)}\rangle_{(x,y)}\quad\text{,}
\end{eqnarray}
where 
\begin{eqnarray}
\mathcal{H}^{(j)}_{\text{eff}} = \Gamma^{\dagger\,(j)} \mathcal{H} \Gamma^{(j)}
\end{eqnarray}
is supposed to be the effective Hamiltonian for the case of Takahashi`s perturbation theory of a specific model in order $j$ perturbation theory. Here $|\psi_1^{(0)}\rangle_{(x,y)}$ corresponds to a one-particle state where the particle is located on site $(x,y)$. Let us mention again that this can be either a spin-flip excitation in a high-field limit or a clusteron excitation which both live on a square lattice.  

The specific $\Gamma \equiv \Gamma^{(\infty)}$ is the operator that transforms an unperturbed state $|\psi_n^{(0)}\rangle$ into the perturbed state space ($\Gamma|\psi_n^{(0)}\rangle = |\psi_n\rangle$). The effective Hamiltonian consists of all possible sequences of creation and annihilation operators, that conserve the total particle number. Particularly for the Hamiltonian $\mathcal{H}$, the effective representation $\mathcal{H}^{(j)}_{\text{eff}}$ allows creation or destruction of $0,1,2$ or $4$ clusterons (see also Fig.~\ref{fig:overview}(b)). Once having computed all possible hopping elements of a given order, one can perform a Fourier transform into momentum space, what leads to the expression for the dispersion
\begin{eqnarray}
\omega(\mathbf{k})^{(j)} = (a^{(j)}_{(0,0)} - E_0^{(j)}) - \sum_{l,m \neq 0} a^{(j)}_{(l,m)}\cos(l k_x + m k_y)\nonumber\quad\text{.}\\
\end{eqnarray}
Finding min($\omega(\mathbf{k})^{(j)}$) leads to the series for the one-particle gap. Let us remark, that the Fourier transform diagonalizes the Hamiltonian in the one-particle sector because the momentum is a good quantum number. Once adding another particle into the system the Hamiltonian will not be diagonal after the Fourier transform, due to the existence of a relative motion.

As a result, we found that the gap is located at momentum $(k_x,k_y)=(\pi,\pi)$ for all limits studied in this work. This allows us to construct a one-particle eigenstate $|\psi_1^{(0)}\rangle_{(k_x,k_y)}$ of the effective Hamiltonian explicitely having $(k_x,k_y)=(\pi,\pi)$. The state $|\psi_1^{(0)}\rangle_{(k_x,k_y)}$ is defined as the Fourier transform of the one-particle states $|\psi_1^{(0)}\rangle_{(x,y)}$ 
\begin{equation}
 |\psi_1^{(0)}\rangle_{(k_x,k_y)} = \frac{1}{N} \sum_{x,y} e^{i \left( k_x x + k_y y\right)}|\psi_1^{(0)}\rangle_{(x,y)} \quad\text{.}
\end{equation}
We therefore have used L\"owdins method to obtain a series expansion with maximal order directly for the one-particle gap.

High-order series expansions for the ground-state energy per site $e_0$ and $\Delta$ have been calculated for different limits of the Hamiltonian \eqref{Mod:hcl}. In particular, we have performed expansions for (i) $J \gg h_x$,$h_z$, (ii) $h_x \gg J$,$h_z$, (iii) and $h_z \gg J$,$h_x$ which are given explicitely in the \ref{Sect:AA}. Note that expansions (i) and (ii) are identical up to an exchange of couplings $J$ and $h_x$ due to the self-duality. Additionally, it is possible to obtain series expansions for $h_x + h_z \gg J$. This is done by applying a base transformation diagonalizing the unperturbed local part $h_x\sum_i\sigma^x_i + h_z\sum_j\sigma^z_j$ of the full Hamiltonian. As a consequence, the latter expansion is the hardest one numerically, because the transformed perturbation (the transformed cluster Hamiltonian) is a very complicated object containing in general all possible five-site matrix elements.

\subsection{iPEPS} 

The method of infinite Projected Entangled Pair States (iPEPS) \cite{Jordan08} produces a variational approximation to the ground-state wavefunction of two-dimensional quantum lattice systems in the thermodynamic limit by employing a tensor-network approach \cite{2004cond.mat..7066V}.  A number of different variations of the method have already been successfully applied to a number of systems \cite{Corboz10, Jordan08, JordanBH, Orus09, CorboztJ, CorbozFrus, BauerSU3, Compass, Dusuel2010, Schulz2012, ctm3d, bela1, bela2, huan, fidPEPS}. In our case, we have adapted the specifics of the algorithm in order to deal efficiently with the peculiarities of the Hamiltonian in Eq.(\ref{hamcl}). Here we explain the most distinctive features of the algorithm that was employed to simulate this model (for generic notions on the method, we address the reader to the afore-mentioned references)

\begin{figure}
\includegraphics[width=0.5\textwidth]{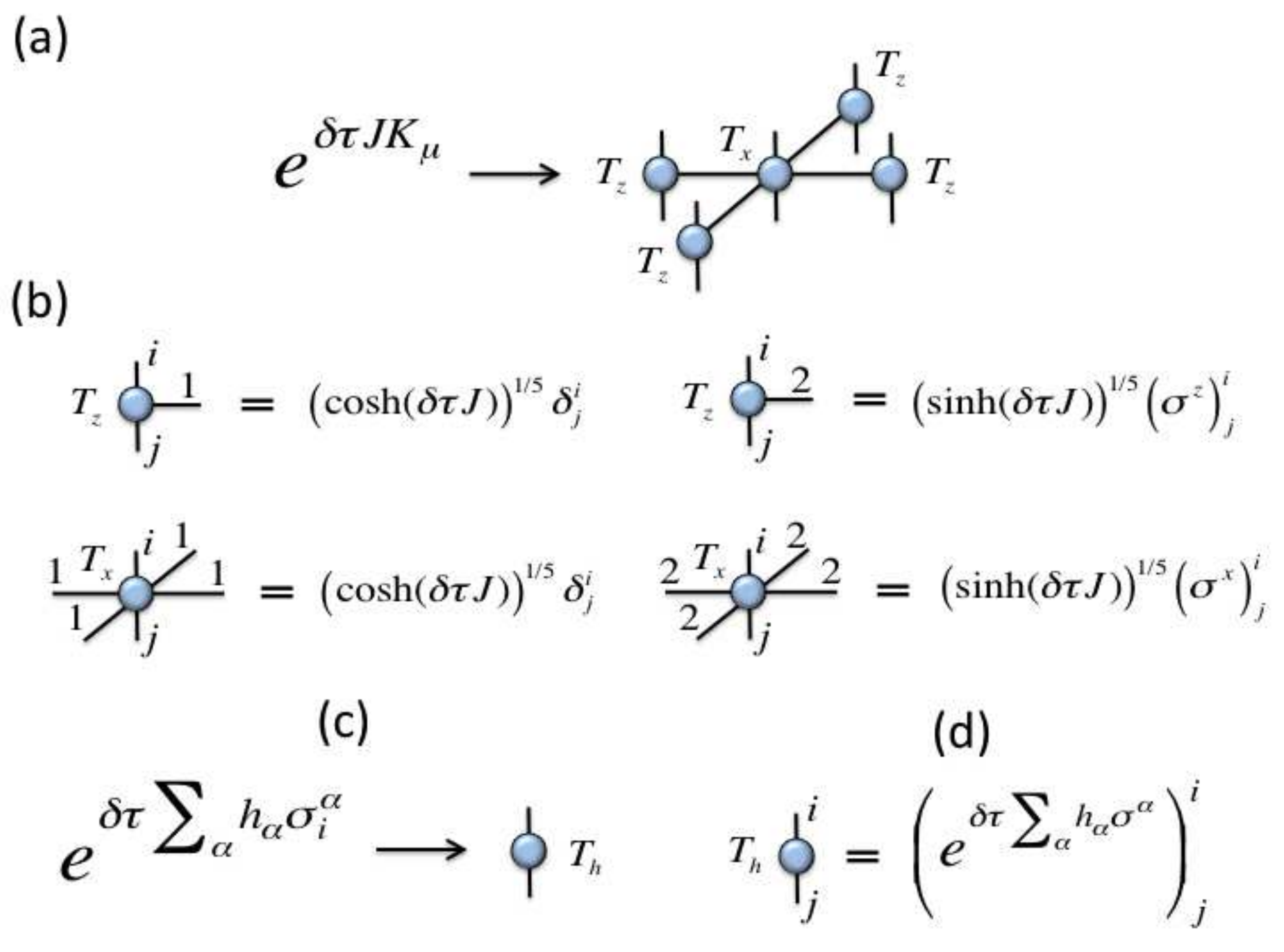}
\caption{(color online) (a) Tensor network diagram for the five-body operator $e^{\delta \tau J K_{\mu}}$. This is given in terms of five tensors, $T_z$ (repeated four times) and $T_x$. (b) Non-zero components of tensors $T_z$ and $T_x$. (c) Tensor network diagram for the one-body operator $e^{\delta \tau \sum_{\alpha} h_{\alpha} \sigma_i^\alpha}$. This is given in terms of one tensor $T_h$. (d) Non-zero components of tensor $T_h$.}
\label{diag1}
\end{figure}

The goal of our algorithm is to best approximate the ground state of $\mathcal{H}$. In general terms, this can be done by implementing, in a way to be specificed later, an imaginary-time evolution driven by the Hamiltonian: 
\begin{equation}
	|\Psi_{\mathrm{gs}}\rangle = \lim_{\tau \rightarrow \infty} \frac{e^{-\tau \mathcal{H} }|\Psi_0\rangle}{||e^{-\tau\mathcal{H} } |\Psi_0\rangle||}\, ,
\end{equation}
where $|\Psi_{\mathrm{gs}}\rangle$ is the ground state of $\mathcal{H}$ and $|\Psi_0\rangle$ is any initial state that has a non-vanishing overlap with the ground state. In order to approximate this evolution, we proceed similarly as explained in e.g. Ref.~\onlinecite{Jordan08,Orus09}. More precisely, the evolution is splitted into small imaginary-time steps $\delta \tau \ll 1$ by using a Suzuki-Trotter expansion of the evolution operator $e^{-\tau \mathcal{H}}$. In our case, for the Hamiltonian in Eq.(\ref{hamcl}) we have
\begin{eqnarray}
	e^{-\tau \mathcal{H}} &\approx& \bigotimes_{\mu} \left(e^{\delta \tau J K_{\mu}}\right)\bigotimes_i \left(e^{\delta \tau \sum_{\alpha} h_{\alpha} \sigma_i^\alpha}\right) + O(\delta \tau ^2) \nonumber \\
	&\approx& \left( U(\delta \tau) \right)^{m} + O(\delta \tau ^2)
	\end{eqnarray}
where a simple first-order approximation has been employed. In the above equation, $m = \tau/ \delta \tau$, and $U(\delta \tau) \equiv e^{- \delta \tau\mathcal{H} }$ is an operator acting over the whole system and implementing the imaginary time evolution for a time step $\delta \tau$. 

It is now convenient for us to switch to the language of tensor network diagrams (see e.g. the introduction in Ref.~\onlinecite{ctm3d}). Using this, we can understand the different elements in the above equation in terms of the diagrams in Fig.~\ref{diag1} and Fig.~\ref{diag2}. In particular, the evolution operator $U(\delta \tau)$ can be easily understood as an infinite Projected Entangled Pair Operator (iPEPO) described by just one tensor, see the diagrams in Fig.~\ref{diag2}.

Using the above representation of the evolution operator in terms of an iPEPO, the iPEPS algorithm applied to our case proceeds as follows:

\begin{figure}
\includegraphics[width=0.5\textwidth]{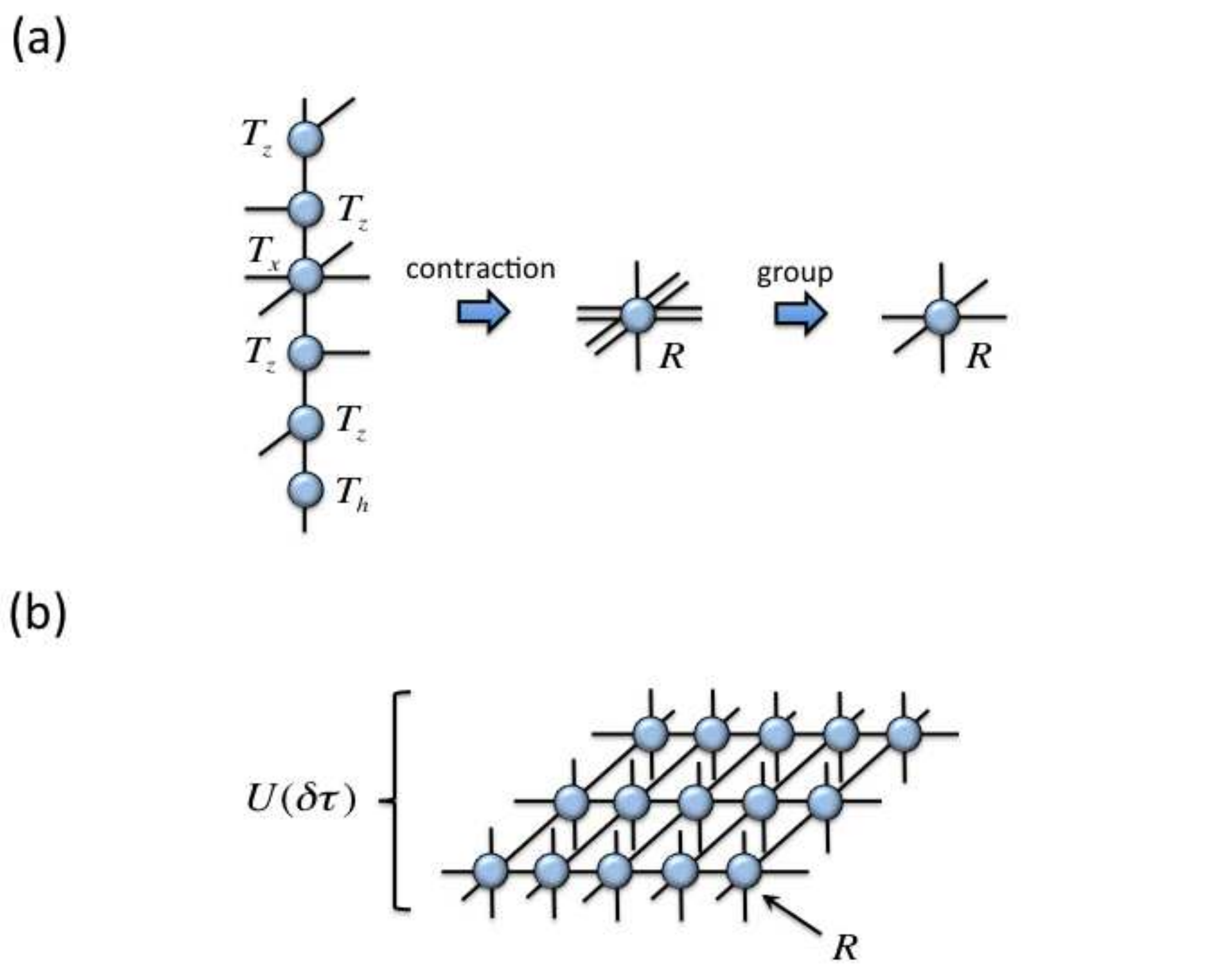}
\caption{(color online) (a) The contraction of four $T_z$ tensors together with $T_x$ and $T_h$ produces a tensor $R$ with double indices, which can be considered as single indices after grouping them. (b) The evolution operator $U(\delta \tau)$ can be understood as an infinite PEPO that can be constructed just from tensor $R$.}
\label{diag2}
\end{figure}

\vspace{8pt} 

\begin{figure}
\includegraphics[width=0.5\textwidth]{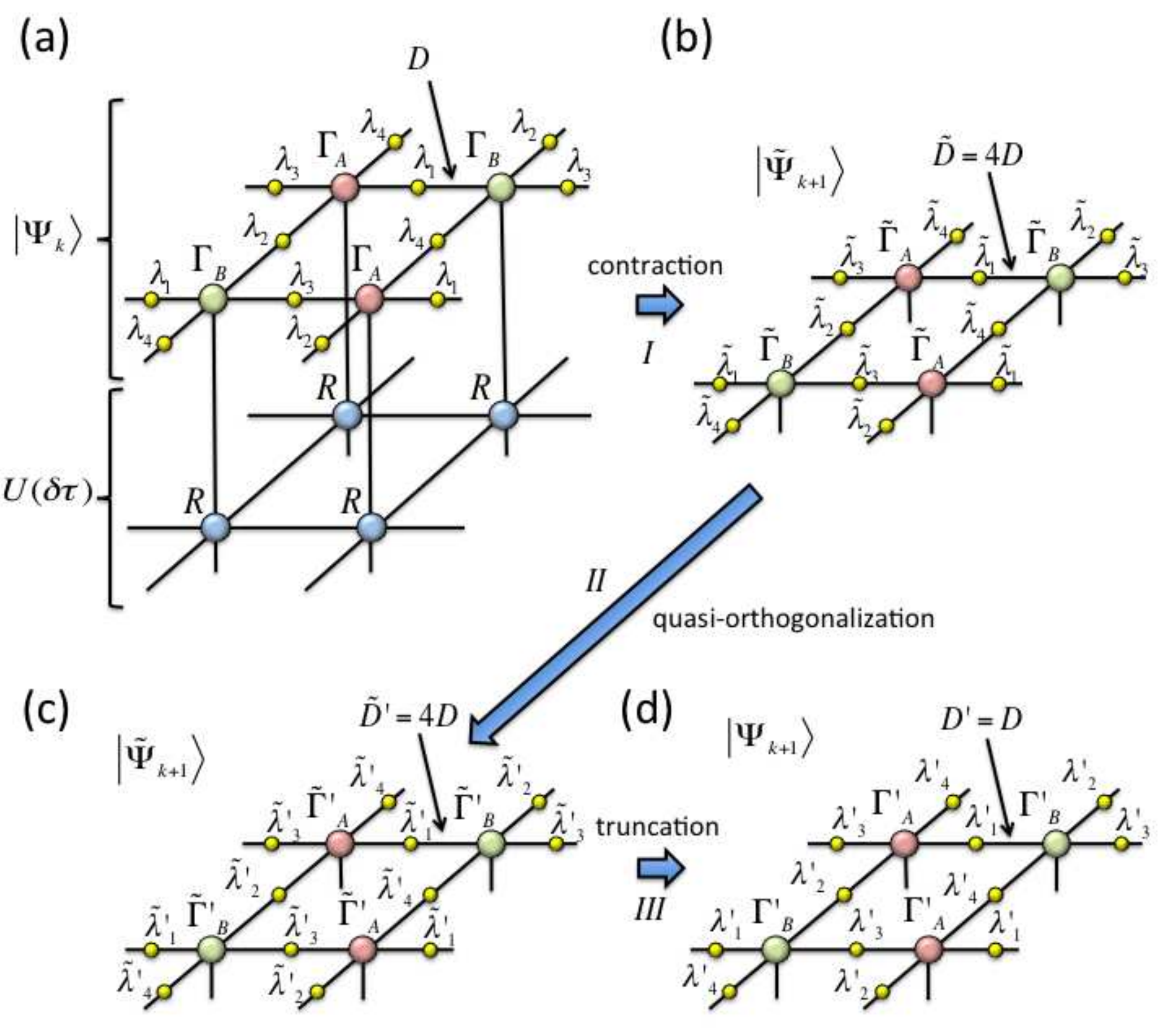}
\caption{(color online) At step $k$, (a) the infinite PEPS for state $\ket{\Psi_k}$ is defined by tensors $\Gamma_A, \Gamma_B$ and positive diagonal matrices $\lambda_1, ..., \lambda_4$. The infinite PEPO for operator $U(\delta \tau)$ is applied. (b) Infinite PEPS for the evolved state $\ket{\tilde{\Psi}_{k+1}}$. Matrices $\tilde{\Gamma}_A, \tilde{\Gamma}_B$ are obtained by contracting ${\Gamma}_A$ and ${\Gamma}_B$ with $R$, whereas matrices $\tilde{\lambda_1},...\tilde{\lambda}_4$ are obtained by doing the tensor product of ${\lambda_1},...{\lambda}_4$ with the $4 \times 4$ identity operator. (c) Quasi-canonical form for the infinite PEPS of state $\ket{\tilde{\Psi}_{k+1}}$. (d) Infinite PEPS for state $\ket{{\Psi}_{k+1}}$.}
\label{diag3}
\end{figure}

(i) \emph{Initialization}: take the initial state $\ket{\Psi_0}$ to be an infinite PEPS with a unit cell of two sites and bond dimension $D$, as in Fig.~\ref{diag3}(a). This is defined in terms of tensors $\Gamma_A, \Gamma_B$ at the lattice sites and the diagonal and positive matrices $\lambda_1, \ldots, \lambda_4$ at the links. This representation for the infinite PEPS is useful in the context of the so-called \emph{simplified update}, see Ref.~\onlinecite{huan, Corboz10}. The initialization can be done in a variety of ways, e.g. by choosing a PEPS that corresponds to the non-perturbed cluster state \cite{PEPSClus}, or either with polarized and random states. 

\vspace{8pt} 
Then, at step $k$, apply the following: 
\vspace{8pt}

(ii) \emph{Contraction}: apply the infinite PEPO for $U(\delta \tau)$ over the infinite PEPS for state $\ket{\Psi_k}$, as in Fig.~\ref{diag3}(a). As a result, obtain a new PEPS of bond dimension $\tilde{D} = 4D$ for the evolved state $\ket{\tilde{\Psi}_{k+1}}$, defined in terms of tensors $\tilde{\Gamma}_A, \tilde{\Gamma}_B$ and matrices $\tilde{\lambda}_1, \ldots, \tilde{\lambda}_4$, see Fig.~\ref{diag3}(b). The complexity of this step is $O(D^4)$. 

\vspace{8pt}

(iii) \emph{Quasi-orthogonalization}: obtain a \emph{quasi canonical form} for the evolved infinite PEPS of state $\ket{\tilde{\Psi}_{k+1}}$. We do this by (a) applying the identity operator over all the links of the lattice, (b) performing simplified updates \cite{huan, Corboz10} to account for the action of these (identity) operators, and (c) iterating this procedure until convergence of the diagonal positive $\lambda$ matrices at each link. The infinite PEPS obtained in this way is defined by tensors $\tilde{\Gamma}'_A, \tilde{\Gamma}'_B$, matrices $\tilde{\lambda}'_1, \ldots, \tilde{\lambda}'_4$, and is reminiscent of the canonical form for infinite Matrix Product States \cite{can2,can3}. Notice, though, that no canonical form exists formally in tensor networks with closed loops such as two-dimensional PEPS. Nevertheless, we expect this procedure to converge quickly for systems with a finite correlation length, and to produce a representation of the infinite PEPS that is well-suited for further numerical manipulations. In practice, we observe that for a wide variety of interesting systems (including the Hamiltonian in Eq.(\ref{hamcl})) this strategy converges very fast numerically. The bond dimension of the resultant infinite PEPS does not change, that is $\tilde{D}' = \tilde{D} = 4D$, see Fig.~\ref{diag3}(c). The complexity of this step is $O(D^5)$. 

\vspace{8pt} 

(iv) \emph{Truncation}: truncate the bond dimension of the infinite PEPS down to $D' = D$ by keeping the largest diagonal elements of the $\lambda$ matrices at each link. The new infinite PEPS for the new state $\ket{{\Psi}_{k+1}}$ is defined by tensors ${\Gamma}'_a, {\Gamma}'_B$ and matrices ${\lambda}'_1, \ldots, {\lambda}'_4$, see Fig.~\ref{diag3}(d).

\vspace{8pt}

(v) Iterate the above procedure for $k=0,1,\ldots$ by applying the infinite PEPO for $U(\delta \tau)$ until the desired convergence has been achieved (in e.g. relevant observables and $\lambda$ matrices). 

\vspace{8pt} 

The above procedure is very similar to the simplified update for two-body gates explained in Ref.~\onlinecite{huan, Corboz10}. Here, though, we use the full power of the infinite PEPO to handle with the five-body interactions in the Hamiltonian in a simple and elegant way. Once convergence has been achieved, we extract expectation values by using e.g. the directional-CTM method explained in Ref.~\onlinecite{Orus09}.

For the purposes of this paper we have seen that an infinite PEPS with bond dimension $2 \le D \le 4$ is already sufficient to produce reliable accuracies in all the results, in combination with a time step $\delta \tau = 10^{-4}$ (relative error in the energy per site of $10^{-3} - 10^{-4}$). Moreover, refinements of the above procedure are also possible by using "full" variational updates of the tensors of the infinite PEPS \cite{Jordan08}. Nevertheless, we have also implemented a number of simulations of the Hamiltonian in Eq.~\eqref{hamcl} using such a "full" variational tensor update, and saw almost no difference in the accuracies of the results.

\subsection{Series expansion plus iPEPS} 

In the following we apply a combined series expansion plus iPEPS approach in order to determine the phase diagram of the perturbed cluster Hamiltonian as it has already been done successfully in the context of perturbed topologically-ordered states \cite{Dusuel2010,Schulz2012}. 

\begin{figure}
\begin{centering}
\includegraphics[width=\columnwidth]{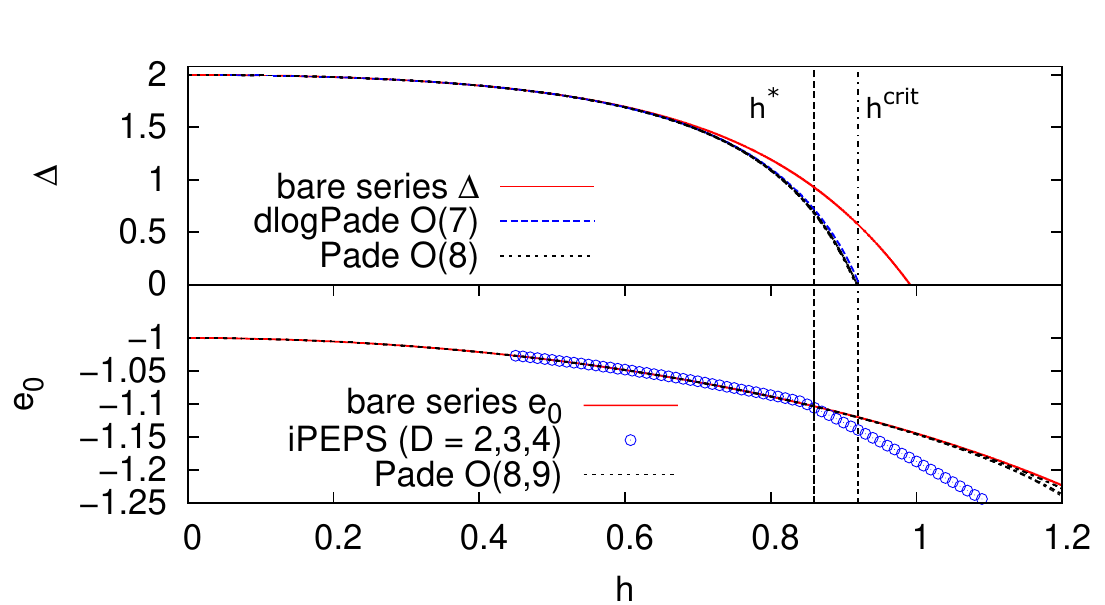} 
\par\end{centering}
\caption{(color online) One-particle gap $\Delta$ (upper panel) and ground-state energy per site $e_0$ (lower panel) as a function of the magnetic field $h$. The field direction is parametrized by $h_x = h \cos(\theta)$ and $h_z = h \sin(\theta)$ and is chosen to be  $\theta = \frac{5}{128} \pi$. Results of series expansion are shown as lines while circles correspond to iPEPS data with bond dimension $D=2,3,4$ (energy differences are negligible in the scale of the plot). The bare series obtained in order 9 (order 8) for $e_0$ ($\Delta$) is plotted as solid red lines. Different dlogPad\'{e} (Pad\'{e}) extrapolations of the highest order are shown as dashed (dotted) lines (deviant extrapolations are omitted in the plot for sake of clarity). Notice that these extrapolations almost collapse in most of the plot. It can be clearly seen that $e_0^{\rm iPEPS}$ is well below $e_0^{\rm SE}$ for a field value for which the one-particle gap is still finite, i.e. $h^{\rm crit}>h^*$ (black vertical lines) holds. We therefore detect a first-order phase transition for $h\approx 0.86$ illustrated as the dashed vertical line at $h^{*}$.
\label{fig:e0vsgap}}
\end{figure}

The underlying physical idea is the following. High-order series expansions of the one-particle gap (or more generally other modes) allows the location of second-order phase transition points. The critical field value $h^{\rm crit}$ corresponds to the field where the one-particle gap vanishes $\Delta(h^{\rm crit})=0$ which often can be determined accurately by resummation techniques like dlogPad\'{e} (Pad\'{e}) extrapolations. In contrast, any series expansion restricted to one limit is not able to detect first-order phase transitions. We therefore define the field $h^*$ for which $e_0^{\rm iPEPS} (h)<e_0^{\rm SE} (h)$ with $h>h^*$ holds. The order of a phase transition is now assigned as follows: If $h^{\rm crit}<h^*$, we detect a second-order phase transition at $h^{\rm crit}$. If $h^{\rm crit}>h^*$, we detect a first-order phase transition at $h^*$ because the series expansion has missed a level crossing in the ground state observed in the variational iPEPS calculation. A typical example for the current problem of the perturbed cluster Hamiltonian is shown in Fig.~\ref{fig:e0vsgap}.

\section{Phase diagram}
\label{Sect:PD}
In this section we present our results for the zero-temperature phase diagram of the cluster Hamiltonian in the presence of external fields which we have obtained by the combined series expansion plus iPEPS approach introduced in the last section. 

In order to simplify the presentation of the phase diagram, we use a coordinate transformation that maps the three-dimensional parameter space spanned by the basis vectors $\{(J,0,0)^T,(0,h_z,0)^T,(0,0,h_x)^T\}$) onto a two-dimensional triangle. Explicitely, the transformation reads
\begin{eqnarray}
  X &=& \frac{1}{\sqrt{2}}\left(1-J+h_x\right)  \nonumber \\
  Y &=& \frac{1}{\sqrt{2}}\left(1-J-h_x\right) \quad ,
\end{eqnarray}
where the normalization is chosen to be $ h_x + h_z + J = 1$. Let us note that we will nevertheless refer to certain points in parameter space using the physically more intuitive three-dimensional coordinates $(J,h_z,h_x)$. The final phase diagram obtained by the series expansion plus iPEPS approach is shown in Fig.~\ref{fig:PD}.  

The phase diagram is symmetric about the centerline of the triangle (solid red line in Fig.~\ref{fig:PD}) which is a direct consequence of the self-duality. The centerline is in fact the self-dual line. The left edge (right edge) of the triangle corresponds to the exactly solvable case $(J,h_z,0)$ ($(0,h_z,h_x)$) already discussed in Sect.~\ref{Sect:Model}. Here the system shows no phase transitions. Finally, the baseline of the triangle represents the model $(J,0,h_x)$ which displays a strong first-order phase transition at $h_x=J$ (lower end point of the red line in Fig.~\ref{fig:PD})). Altogether, it is therefore possible to adiabatically connect all points on the edge of the triangle without encountering any phase transition. 

For the general case of $J$, $h_x$, and $h_z$ finite, we have to use the series expansion plus iPEPS approach about two different limits to deduce the full phase diagram displayed in Fig.~\ref{fig:PD}. 

First, we compared series expansions and iPEPS data in the cluster phase, i.e.~high-order series expansions about the limit $J\gg h_x,h_z$ are performed. A convenient parameterization for the obtained series is to set $h_x = h\cos(\theta)$ and $h_z = h\sin(\theta)$ for $J=1$ and to compare for different values of $\theta$ series expansion and iPEPS data. One typical example is displayed in Fig.~\ref{fig:e0vsgap} and has already been discussed in the last section. We find two (symmetric) first-order lines emerging out of the already known first-order self-dual point at $h_z=0$. The difference between $h^*$ and $h^{\rm crit}$ becomes smaller and smaller when increasing $\theta$ signaling a weakening of the first-order nature of the transition. At a certain point (indicated as blue circles in Fig.~\ref{fig:PD}) $h^*$ and $h^{\rm crit}$ are comparable consistent with a critical end point of the first-order transition lines. Increasing the angle $\theta$ to even larger values, no transition at all is detected.
      
\begin{figure}
\begin{centering}
\includegraphics[width=1.0\columnwidth]{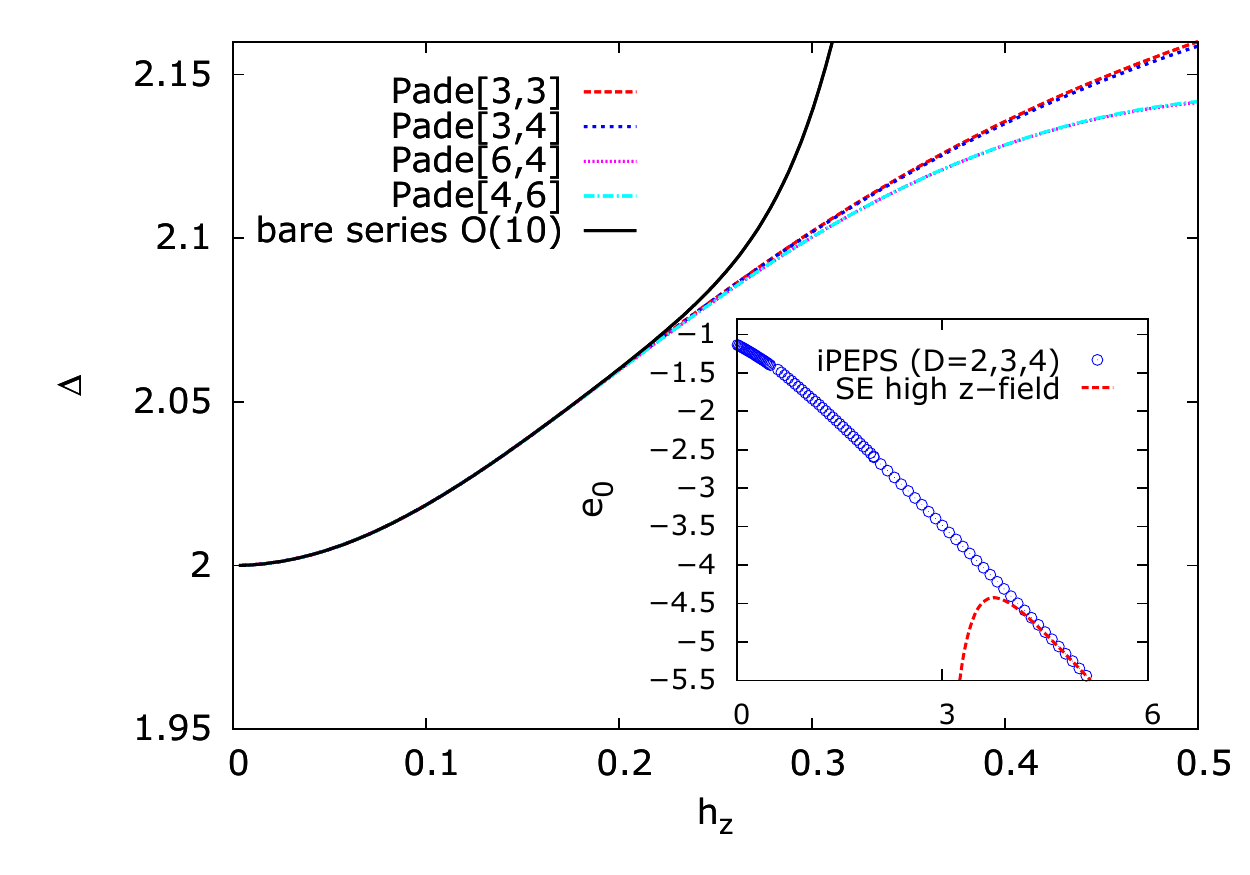} 
\par\end{centering}
\caption{(color online) The gap $\Delta/J$ as a function of $h_z$ is shown along the self dual line $h_x = J$ setting $J=1$. Solid line corresponds to bare series and dashed lines represent different Pad\'{e} extrapolations. We remark that dlogPad\'{e} extrapolations reveal no poles along the real axis. (inset) Comparing the iPEPS data with the series expansion for the ground-state energy per site $e_0$ in the high $h_z$-field case. In this limit $e_0$ behaves linearly with the field $e_0/J=-h_z$. The good agreement between the different limits indicates that there is no phase transition along the full self-dual line.}
\label{fig:SDL}
\end{figure}

Second, we analyzed our data coming from the limit $h_z\gg h_x,J$. Here we find that the one-particle gap shows no tendency to close (in fact it increases) for any combination of $J$ and $h_x$. This is consistent with the absence of any second-order phase transition at least in the convergence radius of our series expansion about the limit $h_z\gg h_x,J$. Furthermore, putting series expansion and iPEPS together we find no evidence for additional phase transition lines except the two first-order phase transition lines already discussed in the last paragraph. As a typical example confirming this scenario, we show in Fig.~\ref{fig:SDL} our data on the self-dual line. Clearly, no evidence of an additional phase transition can be seen. 
    
\begin{figure}
\begin{centering}
\includegraphics[width=1.0\columnwidth,angle=0]{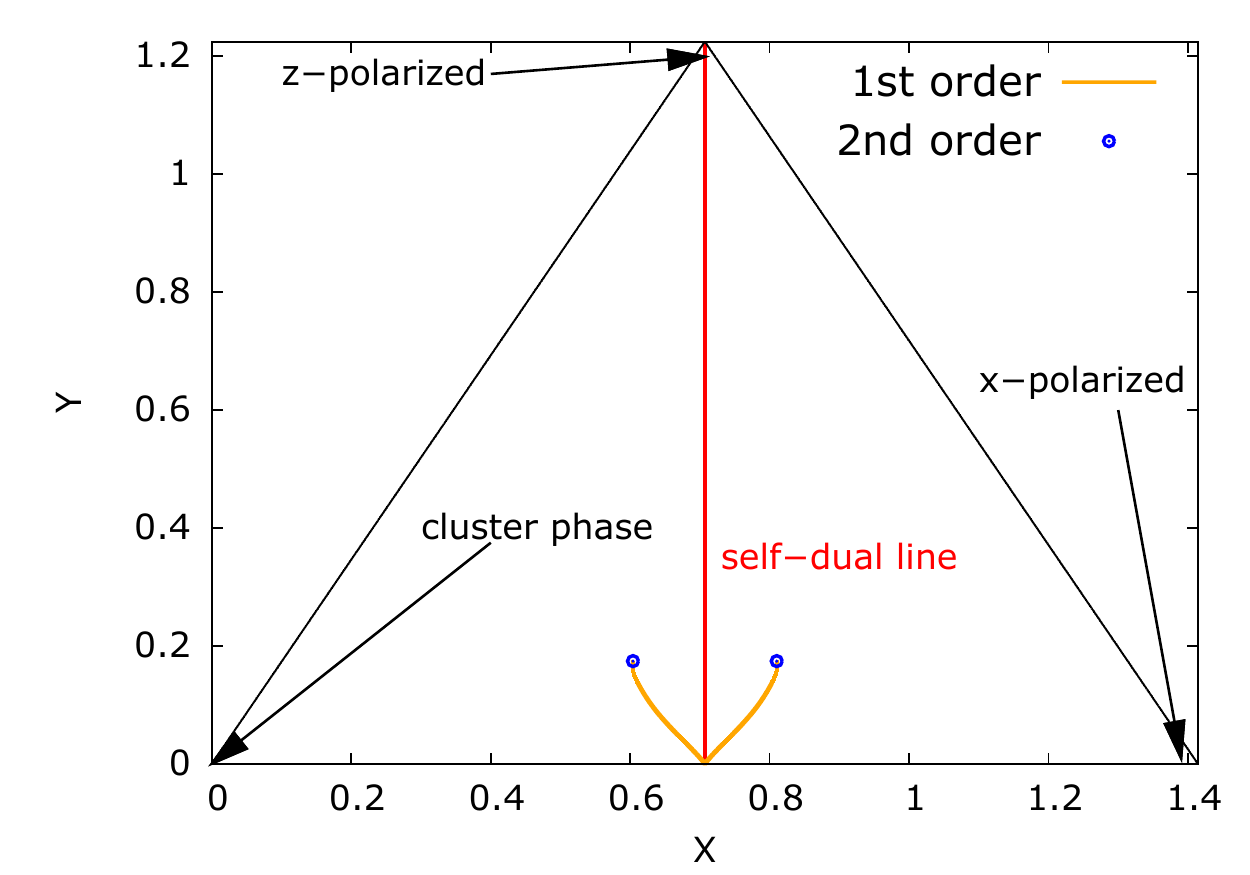} 
\par\end{centering}
\caption{(color online) Phase diagram of the perturbed cluster Hamiltonian as a function of $X$ and $Y$ obtained by the series expansion plus iPEPS approach. The corners of the triangle correspond to the three limits (i) pure cluster Hamiltonian $(J,0,0)$ (left corner), (ii) pure $h_x$-field $(0,0,h_x)$ (right corner), and (iii) pure $h_z$-field $(0,h_z,0)$ (upper corner). The self-dual line is illustrated by the vertical red solid line cutting the triangle in the middle. The whole spectrum and therefore also the phase diagram on both sides of the self-dual line is fully symmetric. First-order phase transition lines are drawn as orange solid lines. The two critical end points are marked by blue circles (coordinates of the left point $(X_{\rm left} \approx 0.604, Y_{\rm left} \approx 0.174)$ and coordinates of the right point $(X_{\rm right} \approx 0.810, Y_{\rm right} = Y_{\rm left})$).\label{fig:PD}}
\label{fig:PD}
\end{figure}

\section{Fidelity}
\label{Sect:F}
From the condensed matter point of view, the main questions are addressed by determining the zero-temperature phase diagram. However, the phase boundaries only correspond to an upper bound for the usability in measurement-based quantum computing. In the following we want to pinpoint the boundaries quantitatively for which the entanglement properties of a perturbed cluster state at finite fields is still appropriate for measurement-based quantum computing. In order to answer this question, we calculate the fidelity per site $d$ of the perturbed cluster state at finite fields with the exact cluster state measuring the distance between two quantum states. In fact, it has been shown that the cluster phase is still usable for measurement-based quantum computing by applying quantum error correction techniques when the fidelity per site $d$ is larger than $0.986$ \cite{Raussendorf20062242}.

Let us remark, that the fidelity is not a metric on density operators, even though it is used as a  measure to estimate distances. In this paper we evaluate only the fidelity between pure states. Nevertheless, in what follows we give a short general introduction to this quantity.

\begin{figure}
\begin{centering}
\includegraphics[width=1.0\columnwidth]{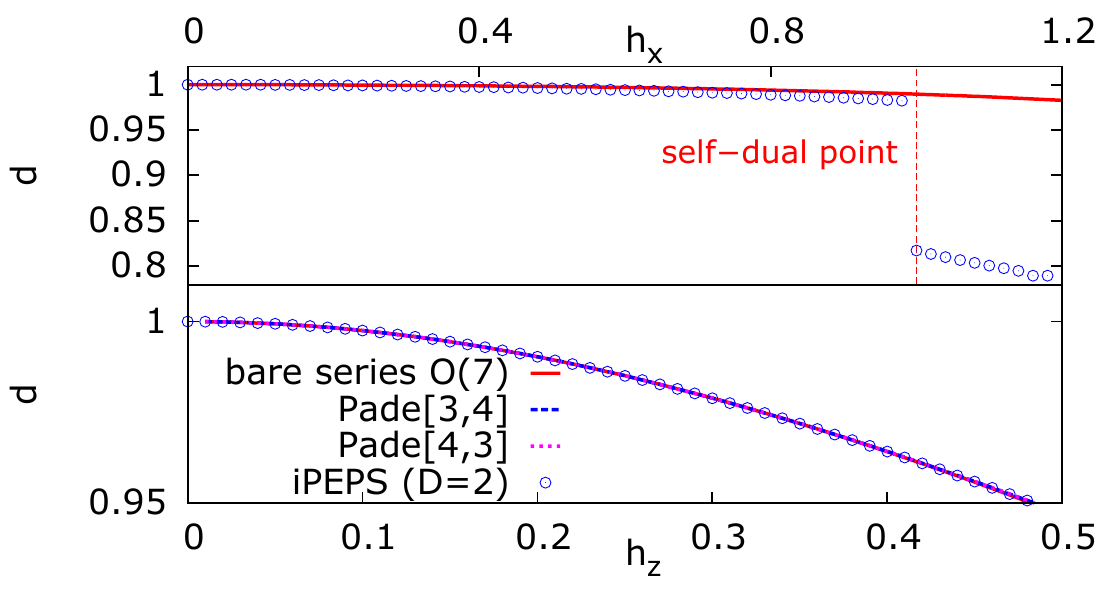} 
\par\end{centering}
\caption{(color online) Direct comparison between the series expansion (SE) for the fidelity and the iPEPS results. We contrast the results in the limit $h_z = 0$ (upper panel) and $h_x = 0$ (lower panel). One clearly sees the deviation of both approaches near the first order transition (red dashed line). We acribe this fact to the insensitivity of the SE-approach to first order effects. The agreement along the $h_z$-axis remains very good, considering the Pad\'{e} extrapolations of the bare series. Notice also the collapse of the different Pad\'e extrapolations with the bare series in the scale of the plot. \label{fig:fid_comp}}
\end{figure}

\begin{figure}
\begin{centering}
\includegraphics[width=1.0\columnwidth]{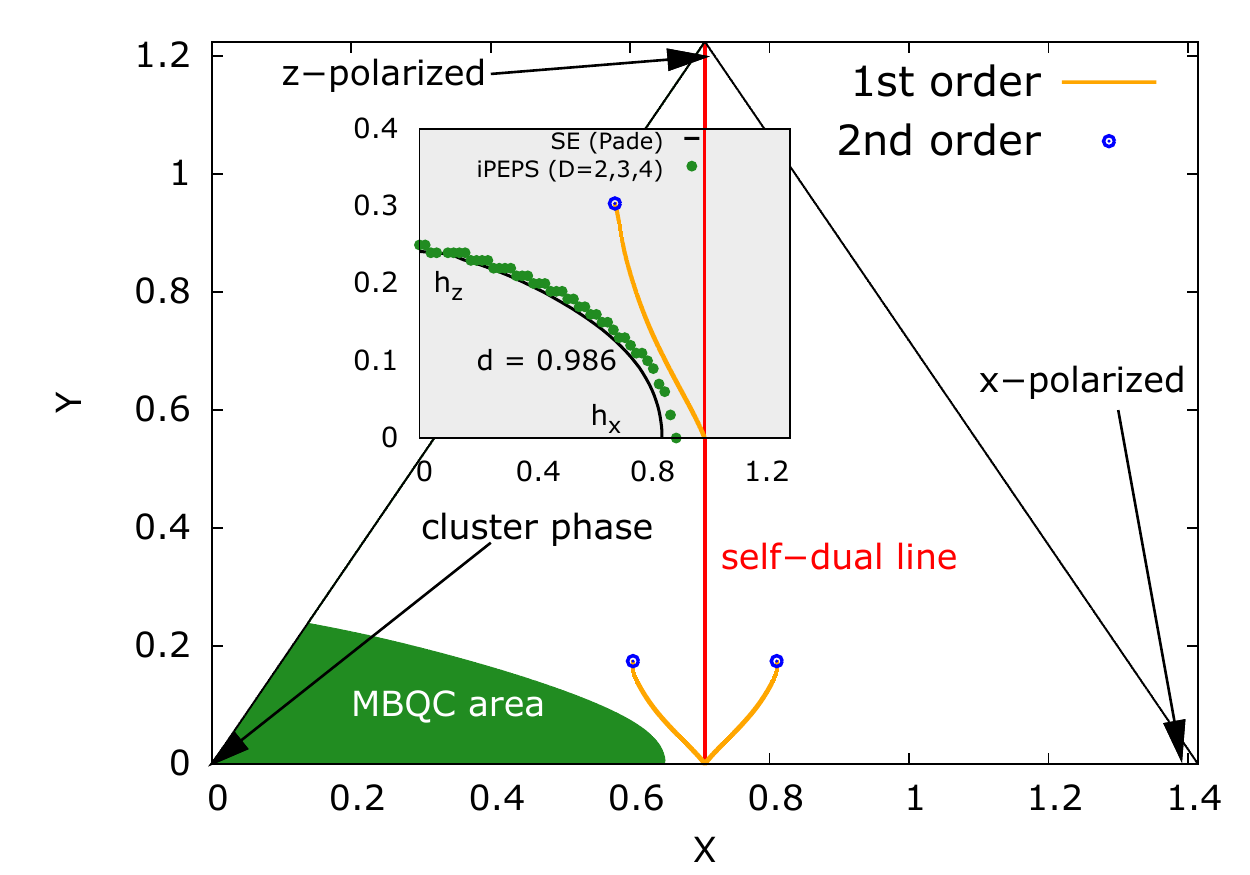} 
\par\end{centering}
\caption{(color online) Phase diagram including the useful region for quantum computing. The shaded grey (green) area and the grey inset indicate the usable region for measurement based quantum computing (MBQC) according to the fidelity threshold at $d = 0.986$ \cite{Raussendorf20062242}. (inset) Comparing the extrapolations for the series expansion (SE) of the fidelity per site (black line) with the iPEPS results (green dots). Clearly there is no agreement beyond the first order transition line ($h_x > h^*$). This is an artifact of the series expansion technique, which is not sensitive to first order effects. We emphasize this fact by plotting the extrapolations beyond this point with a dashed line. Departing from the phase boundary, both fidelities are in very good agreement.\label{fig:fid}}
\end{figure}

The fidelity for two quantum states governed by their density matrices $\rho$ and $\sigma$ is defined as
\begin{eqnarray}
F(\rho,\sigma) \equiv \text{tr}\left(\sqrt{\sqrt{\rho}\sigma\sqrt{\rho}}\right) \label{fidelitydef} \quad .
\end{eqnarray}
Let $\rho = \sum_xp_x|x\rangle\langle x|$ and $\sigma = \sum_xq_x|x\rangle\langle x|$ be two commuting density matrices, i.e. they can be diagonalized in the same orthogonal basis. One can show that definition \eqref{fidelitydef} will reproduce the defintion of the fidelity in classical probability theory
\begin{eqnarray}
F(\rho,\sigma) = F(p_x,q_x) = \sum_x\sqrt{p_xq_x} \quad .
\end{eqnarray}
Later we will explicitely use the invariance of the fidelity under unitary transformations. Using Uhlmann`s theorem one finds that the fidelity has the characteristics of an overlap of two wavefunctions
\begin{eqnarray}
F(\rho,\sigma) = \max_{|\psi\rangle,|\phi\rangle} |\langle\psi|\phi\rangle| \quad ,
\end{eqnarray}
where $|\psi\rangle,|\phi\rangle$ are purifications of the respective density operators. This general definition reduces for the case of one pure state $\rho = |\phi\rangle\langle\phi|$ and an arbitrary (mixed) state $\sigma$ to
\begin{eqnarray}
F(\rho,\sigma) = \text{tr}\left(\sqrt{\langle\phi|\sigma|\phi\rangle|\phi\rangle\langle\phi|}\right) = \sqrt{\langle\phi|\sigma|\phi\rangle} \quad .
\end{eqnarray}
The square of $F(\rho,\sigma)$ corresponds to the probability of finding $\sigma$ in the pure state $|\phi\rangle$. Here we need the fidelity of two pure states $\rho = |\phi\rangle\langle\phi|$ and $\sigma = |\psi\rangle\langle\psi|$ which is given by
\begin{eqnarray}
F(\rho,\sigma) = |\langle\phi|\psi\rangle| \quad .
\end{eqnarray}
It has been shown \cite{1751-8121-41-41-412001} that the fideltiy $F(\rho(\lambda),\rho(\lambda'))$ of a quantum system $H = H_0 + \lambda V$ is an extensive quantity which scales exponentially with the number of lattice sites $N$. Since our aim is to quantify the system behaviour in the thermodynamic limit we use the fidelity per lattice site $d$ which can be derived from the fidelity as follows
\begin{eqnarray}
d = \lim_{N\rightarrow \infty}\sqrt[N]{F(\rho(\lambda),\rho(\lambda'))^2} \quad .
\end{eqnarray}
We calculated the fidelity per site $d$ between the ground state of the system at finite magnetic field and the unperturbed cluster state, both using iPEPS and high-order series expansions. In the context of iPEPS, this quantity can be easily computed by using e.g. the techniques explained in Ref.~\onlinecite{fidPEPS} (and generalizations thereof). With series expansions, it can be obtained using Takahashi`s degenerate perturbation theory \cite{0022-3719-10-8-031} as discussed recently \cite{arXiv:1111.5945v1}. The explicit expression of $d$ reads 
\begin{eqnarray}
d = \lim_{N\rightarrow \infty}\sqrt[N]{|\langle \psi_{\text{CS}}|\Gamma|\psi_{\text{CS}} \rangle|^2} \quad , \label{fidpersite}
\end{eqnarray}
where the unperturbed ground-state wavefunction is set to the exact cluster state $|\psi_0^{(0)}\rangle = |\psi_{\text{CS}} \rangle$. Let us remark, that expression \eqref{fidpersite} suggests $\Gamma$ to be independent of the order perturbation theory $j$, but in fact $\Gamma$ is also an operator sequence of a certain order. In general $\Gamma$ transforms the unperturbed state into the perturbed subspace (see also Sect.~\ref{Sect:Methods}). Explicitely the series for $d$ is given in the \ref{Sect:AC}. Comparing the series expansion approach with the iPEPS data (see Fig.~\ref{fig:fid_comp}) one finds a very good agreement of the two approaches except for the first-order line. Let us stress that by definition the series expansion cannot capture the jump of the fidelity at the transition. Due to this fact the iPEPS data is the only reliable tool to study the fidelity beyond the phase boundary. In Fig.~\ref{fig:fid} we present the phase diagram in combination with the results for the fidelity threshold $d > 0.986$. Clearly, almost the whole cluster phase up to the first-order phase transition line lies above the usability threshold, what is indeed a promising feature for measurement-based quantum computing.

\section{Conclusions}
\label{Sect:Conclusion}
In this work we have studied the influence of an external magnetic field on the so-called cluster 
state being a highly-entangled state relevant for measurement-based quantum computing. Concretely, 
this is done by analysing the cluster Hamiltonian in the presence of external magnetic fields $h_x$
 and $h_z$ on the two-dimensional square lattice using a combination of high-order series expansions and variational iPEPS 
calculations.

We find an interesting zero-temperature phase diagram displaying the cluster phase and polarized phases.
 The phase diagram is fully symmetric under the exchange of $J$ and $h_x$ due to the existence of a self-dual line
 in parameter space. Furthermore we showed, that the self-duality also holds true for the cluster Hamiltonian in combination with a 
$h_y$-field. The phase diagram is dominated by two first-order lines related by self-duality which emerge out
 of the self-dual point for $h_z=0$. The end points of both lines for finite $h_z$ are critical. Unfortunately, our 
current data does not allow to determine the critical exponents and therefore to pinpoint the universality class of these
 critical end points. Additionally, our results show that all ground states appearing in the full parameter space 
can be connected adiabatically.

From the quantum information point of view, the main interest in studying a perturbed cluster Hamiltonian is to quantify
 the robustness and the usefulness of the perturbed cluster state for measurement-based quantum computing. To this end we
 have calculated the fidelity per site of the perturbed ground state at finite fields with the exact cluster state. 
We found that the fidelity per site remains remarkably high in a large part of the parameter space. This is a direct 
 consequence of the fact that our phase diagram is dominated by first-order phase transitions.  

Finally, we would like to remark that most of the qualitative aspects of our study are also true for other lattice topologies.
This is a consequence of two points. First, the self-dual line exists on any lattice. Second, the case of a single field in 
$z$-direction remains always exactly solvable giving no phase transition. Both facts constrain the shape of the phase diagram
 and its properties on most lattice topologies. The situation can be different if other types of external perturbations 
are present which possibly lead to second-order phase transitions. 
   
K.P.S. acknowledges ESF and EuroHorcs for funding through his EURYI. 
R.O. acknowledges financial support from the EU through a Marie Curie International Incoming Fellowship.

\bibliography{./literatur}
\onecolumngrid
\setcounter{section}{0}
\renewcommand*\thesection{Appendix \Alph{section}}
\renewcommand*\thesubsection{\Roman{subsection}}
\section{\label{Sect:A}}
\subsection {Series expansion $h_z \gg J,h_x$ }
\label{Sect:AA}
By exploring the model in the limit $h_z \gg J,h_x$ (fixing the energy scale to $h_z = 1$), one can do a perturbative expansion about the z-polarized phase. Let us shortly remark, that quasi particles in this limit are considered to be simple spin flips (magnons). We obtain the following series expansion for the ground-state energy per site up to order $14$ using L\"owdin's approach: 
\begin{eqnarray*}
e_0^{(14)} &=& -1-\frac{1}{2}\,{{\it h_x}}^{2}-\frac{1}{2}\,{J}^{2}+\frac{1}{8}\,{J}^{4}+\frac{1}{8}\,{{\it h_x}}^{4}-\frac{1}{16}\,{J}^{6}-\frac{1}{16}\,{{\it h_x}}^{6}+{\frac {5}{128}}\,{J}^{8}\\
&&+{\frac {5}{128}}\,{{\it h_x}}^{8}-{\frac {7}{256}}\,{J}^{10}-{\frac {7}{256}}\,{{\it h_x}}^{10}+5/2\,{\it h_x}\,{J}^{3}+{\frac {19}{4}}\,{{\it h_x}}^{2}{J}^{2}+5/2\,{{\it h_x}}^{3}J-{\frac {197}{4}}\,{{\it h_x}}^{3}{J}^{3}\\
&&-{\frac {463}{16}}\,{{\it h_x}}^{4}{J}^{2}-{\frac {35}{8}}\,{{\it h_x}}^{5}J-{\frac {35}{8}}\,{\it h_x}\,{J}^{5}-{\frac {463}{16}}\,{{\it h_x}}^{2}{J}^{4}+{\frac {105}{16}}\,{\it h_x}\,{J}^{7}+{\frac {3347}{32}}\,{{\it h_x}}^{2}{J}^{6}\\
&&+{\frac {62935}{144}}\,{{\it h_x}}^{3}{J}^{5}+{\frac {390503}{576}}\,{{\it h_x}}^{4}{J}^{4}+{\frac {62935}{144}}\,{{\it h_x}}^{5}{J}^{3}+{\frac {3347}{32}}\,{{\it h_x}}^{6}{J}^{2}+{\frac {105}{16}}\,{{\it h_x}}^{7}J\\
&&-{\frac {1155}{128}}\,{\it h_x}\,{J}^{9}-{\frac {1082869}{3840}}\,{{\it h_x}}^{2}{J}^{8}-{\frac {75441}{32}}\,{{\it h_x}}^{3}{J}^{7}-{\frac {130301657}{17280}}\,{{\it h_x}}^{4}{J}^{6}-{\frac {18856195}{1728}}\,{{\it h_x}}^{5}{J}^{5}\\
&&-{\frac {130301657}{17280}}\,{{\it h_x}}^{6}{J}^{4}-{\frac {75441}{32}}\,{{\it h_x}}^{7}{J}^{3}-{\frac {1082869}{3840}}\,{{\it h_x}}^{8}{J}^{2}-{\frac {1155}{128}}\,{{\it h_x}}^{9}J\\
&&+{\frac {3003}{256}}\,{\it h_x}\,{J}^{11}+{\frac {4859819}{7680}}\,{{\it h_x}}^{2}{J}^{10}+{\frac {11847547}{1280}}\,{{\it h_x}}^{3}{J}^{9}+{\frac {12177924859}{230400}}\,{{\it h_x}}^{4}{J}^{8}\\
&&+{\frac {36618487259}{259200}}\,{{\it h_x}}^{5}{J}^{7}+{\frac {100625865563}{518400}}\,{{\it h_x}}^{6}{J}^{6}+{\frac {36618487259}{259200}}\,{{\it h_x}}^{7}{J}^{5}+{\frac {12177924859}{230400}}\,{{\it h_x}}^{8}{J}^{4}\\
&&+{\frac {11847547}{1280}}\,{{\it h_x}}^{9}{J}^{3}+{\frac {4859819}{7680}}\,{{\it h_x}}^{10}{J}^{2}+{\frac {3003}{256}}\,{{\it h_x}}^{11}J-{\frac {15015}{1024}}\,{\it h_x}\,{J}^{13}-{\frac {38471477}{30720}}\,{{\it h_x}}^{2}{J}^{12}\\
&&-{\frac {224644117}{7680}}\,{{\it h_x}}^{3}{J}^{11}-{\frac {1869780914599}{6912000}}\,{{\it h_x}}^{4}{J}^{10}-{\frac {12384846807317}{10368000}}\,{{\it h_x}}^{5}{J}^{9}-{\frac {174347365283677}{62208000}}\,{{\it h_x}}^{6}{J}^{8}\\
&&-{\frac {28780922226379}{7776000}}\,{{\it h_x}}^{7}{J}^{7}-{\frac {174347365283677}{62208000}}\,{{\it h_x}}^{8}{J}^{6}-{\frac {12384846807317}{10368000}}\,{{\it h_x}}^{9}{J}^{5}-{\frac {1869780914599}{6912000}}\,{{\it h_x}}^{10}{J}^{4}\\
&&-{\frac {224644117}{7680}}\,{{\it h_x}}^{11}{J}^{3}-{\frac {38471477}{30720}}\,{{\it h_x}}^{12}{J}^{2}-{\frac {15015}{1024}}\,{{\it h_x}}^{13}J+{\frac {21}{1024}}\,{{\it h_x}}^{12}-{\it h_x}\,J-{\frac {33}{2048}}\,{J}^{14}\\
&&-{\frac {33}{2048}}\,{{\it h_x}}^{14}+{\frac {21}{1024}}\,{J}^{12}\quad \text{.}
\end{eqnarray*}
One directly can confirm this series to be self-dual, under the exchange of $J$ and $h_x$. Furthermore we computed the one-particle gap up to order $10$ using L\"owdin's approach:
\begin{eqnarray*}
\Delta^{(10)} &=& 2+{J}^{2}+2\,{\it h_x}\,J+{{\it h_x}}^{2}-\frac{1}{4}\,{J}^{4}-17\,{\it h_x}\,{J}^{3}-{\frac {67}{2}}\,{{\it h_x}}^{2}{J}^{2}-17\,{{\it h_x}}^{3}J-\frac{1}{4}\,{{\it h_x}}^{4}+\frac{1}{8}\,{J}^{6}\,+{\frac {155}{4}}\,{\it h_x}\,{J}^{5}\\
&&+{\frac {6893}{24}}\,{{\it h_x}}^{2}{J}^{4}+{\frac {2983}{6}}\,{{\it h_x}}^{3}{J}^{3}+{\frac {6893}{24}}\,{{\it h_x}}^{4}{J}^{2}+{\frac {155}{4}}\,{{\it h_x}}^{5}J+\frac{1}{8}\,{{\it h_x}}^{6}-{\frac {5}{64}}\,{J}^{8}\\
&&-{\frac {525}{8}}\,{\it h_x}\,{J}^{7}-{\frac {19499}{16}}\,{{\it h_x}}^{2}{J}^{6}-{\frac {127073}{24}}\,{{\it h_x}}^{3}{J}^{5}-{\frac {795181}{96}}\,{{\it h_x}}^{4}{J}^{4}-{\frac {127073}{24}}\,{{\it h_x}}^{5}{J}^{3}\\
&&-{\frac {19499}{16}}\,{{\it h_x}}^{6}{J}^{2}-{\frac {525}{8}}\,{{\it h_x}}^{7}J-{\frac {5}{64}}\,{{\it h_x}}^{8}+{\frac {7}{128}}\,{J}^{10}+{\frac {6195}{64}}\,{\it h_x}\,{J}^{9}+{\frac {1383121}{384}}\,{{\it h_x}}^{2}{J}^{8}\\
&&+{\frac {22957873}{720}}\,{{\it h_x}}^{3}{J}^{7}+{\frac {99787009}{960}}\,{{\it h_x}}^{4}{J}^{6}+{\frac {43524865}{288}}\,{{\it h_x}}^{5}{J}^{5}+{\frac {99787009}{960}}\,{{\it h_x}}^{6}{J}^{4}\\
&&+{\frac {22957873}{720}}\,{{\it h_x}}^{7}{J}^{3}+{\frac {1383121}{384}}\,{{\it h_x}}^{8}{J}^{2}+{\frac {6195}{64}}\,{{\it h_x}}^{9}J+{\frac {7}{128}}\,{{\it h_x}}^{10} \quad \text{.}
\end{eqnarray*}
\subsection {Series expansion $J \gg h_z,h_x$}
\label{Sect:AB}
Performing the series expansion in the cluster phase, one obtains (fixing the energy scale to $J = 1$) the following expression for the ground-state energy per site in order $9$ using L\"owdin's approach:
\begin{eqnarray*}
e_0^{(9)} &=& -1-\frac{1}{2}\,{{\it h_z}}^{2}+\frac{1}{8}\,{{\it h_z}}^{4}-\frac{1}{16}\,{{\it h_z}}^{6}+{\frac {5}{128}}\,{{\it h_z}}^{8}-{{\it h_z}}^{4}{\it h_x}+\frac{5}{2}\,{{\it h_z}}^{6}{\it h_x}-{\frac {35}{8}}\,{{\it h_z}}^{8}{\it h_x}-\frac{1}{8}\,{{\it h_x}}^{2}\\
&&-{\frac {19}{240}}\,{{\it h_z}}^{2}{{\it h_x}}^{2}-{\frac {977}{960}}\,{{\it h_z}}^{4}{{\it h_x}}^{2}-{\frac {1187}{1920}}\,{{\it h_z}}^{6}{{\it h_x}}^{2}-{\frac {11}{9}}\,{{\it h_z}}^{4}{{\it h_x}}^{3}\\
&&-{\frac {4127}{720}}\,{{\it h_z}}^{6}{{\it h_x}}^{3}-{\frac {13}{1536}}\,{{\it h_x}}^{4}-{\frac {764543}{24192000}}\,{{\it h_z}}^{2}{{\it h_x}}^{4}-{\frac {17378239}{13824000}}\,{{\it h_z}}^{4}{{\it h_x}}^{4}\\
&&-{\frac {690879571}{635040000}}\,{{\it h_z}}^{4}{{\it h_x}}^{5}-{\frac {197}{98304}}\,{{\it h_x}}^{6}-{\frac {334349031161}{20863180800000}}\,{{\it h_z}}^{2}{{\it h_x}}^{6}-{\frac {163885}{226492416}}\,{{\it h_x}}^{8}
\end{eqnarray*}
Furthermore, for the one-particle gap up to order $8$ using L\"owdin's approach one receives:
\begin{eqnarray*}
\Delta^{(8)} &=& 2-12\,{{\it h_z}}^{2}{\it h_x}+32\,{{\it h_z}}^{4}{\it h_x}-{\frac {115}{2}}\,{{\it h_z}}^{6}{\it h_x}-{\frac {223}{12}}\,{{\it h_z}}^{2}{{\it h_x}}^{2}+{\frac {3323}{48}}\,{{\it h_z}}^{4}{{\it h_x}}^{2}\\
&&-{\frac {2711}{96}}\,{{\it h_z}}^{6}{{\it h_x}}^{2}-{\frac {2345}{144}}\,{{\it h_z}}^{2}{{\it h_x}}^{3}+{\frac {766661}{7200}}\,{{\it h_z}}^{4}{{\it h_x}}^{3}-{\frac {25369919}{1209600}}\,{{\it h_z}}^{2}{{\it h_x}}^{4}\\
&&+{\frac {168566773}{691200}}\,{{\it h_z}}^{4}{{\it h_x}}^{4}-{\frac {3313008739}{211680000}}\,{{\it h_z}}^{2}{{\it h_x}}^{5}-{\frac {44922852472229}{2133734400000}}\,{{\it h_z}}^{2}{{\it h_x}}^{6}+{{\it h_z}}^{2}\\
&&-{\frac {5}{64}}\,{{\it h_z}}^{8}-\frac{1}{4}\,{{\it h_z}}^{4}+\frac{1}{8}\,{{\it h_z}}^{6}-\frac{1}{2}\,{{\it h_x}}^{2}-{\frac {15}{128}}\,{{\it h_x}}^{4}-{\frac {575}{12288}}\,{{\it h_x}}^{6}-{\frac {26492351}{1019215872}}\,{{\it h_x}}^{8}
\end{eqnarray*}
\subsection {Series expansion for the fidelity in the limit $J \gg h_z,h_x$ }
Using Takahashi's perturbation theory, the fidelity per site $d$ up to order 7 (fixing the energy scale to $J = 1$) is given by
\label{Sect:AC}
\begin{eqnarray*}
d^{(7)} &=& 1-\frac{1}{4}\,{{\it h_z}}^{2}-{\frac {1}{64}}\,{{\it h_x}}^{2}+\frac{3}{16}\,{{\it h_z}}^{4}-{\frac {5431}{57600}}\,{{\it h_z}}^{2}{{\it h_x}}^{2}-{\frac {137}{36864}}\,{{\it h_x}}^{4}\\
&&-{\frac {269}{192}}\,{{\it h_z}}^{4}{\it h_x}-{\frac {5}{32}}\,{{\it h_z}}^{6}-{\frac {69287}{46080}}\,{{\it h_z}}^{4}{{\it h_x}}^{2}-{\frac {685710229}{13547520000}}\,{{\it h_z}}^{2}{{\it h_x}}^{4}\\
&&-{\frac {20171}{14155776}}\,{{\it h_x}}^{6}+{\frac {19129}{3840}}\,{{\it h_z}}^{6}{\it h_x}-{\frac {5226287}{2419200}}\,{{\it h_z}}^{4}{{\it h_x}}^{3}\quad \text{.}
\end{eqnarray*}
\phantom{abcdefg}

\end{document}